\documentclass[prb,twocolumn,superscriptaddress,a4paper,showkeys,showpacs]{revtex4}
\usepackage{graphicx}
\usepackage{subfigure}
\usepackage{dcolumn}
\usepackage{booktabs}
\usepackage{units}
\usepackage{amsmath}
\usepackage{amsfonts}
\usepackage{amssymb}
\usepackage{bm}
\usepackage{color}
\usepackage[english]{babel}
\usepackage[latin1]{inputenc}
\usepackage{miller}

\usepackage{booktabs}

\begin{document}

\title{Statistical investigations on nitrogen-vacancy centre creation}
\author{D$.$ Antonov}
\email{d.antonov@physik.uni-stuttgart.de}
\author{T$.$ H\"au\ss ermann}
\author{A$.$ Aird}
\affiliation{3$.$ Physikalisches Institut, Universit\"at Stuttgart, Pfaffenwaldring 57, 70569 Stuttgart, Germany}
\author{J$.$ Roth}
\author{H$.$-R$.$ Trebin}
\affiliation{Institut f\"ur Theoretische und Angewandte Physik (ITAP), Universit\"at Stuttgart, Pfaffenwaldring 57, 70569 Stuttgart, Germany}
\author{C$.$ M\"uller}
\author{L$.$ McGuiness}
\author{F$.$ Jelezko}
\affiliation{Universit\"at Ulm, Albert-Einstein-Allee 11, 89081 Ulm, Germany}
\author{T$.$ Yamamoto}
\affiliation{Japan Atomic Energy Agency, Takasaki, Gunma, 370-1292 Japan}
\author{J$.$ Isoya}
\affiliation{Graduate School of Library, Information and Media Studies, University of Tsukuba, 1-2 Kasuga, Tsukuba, Ibaraki 305-8550, Japan}
\author{S$.$ Pezzagna}
\author{J$.$ Meijer}
\affiliation{ Ruhr-Universit\"at Bochum, Universit\"atsstr$.$ 150,  44780 Bochum, Germany}
\author{J$.$ Wrachtrup}
\affiliation{3$.$ Physikalisches Institut, Universit\"at Stuttgart, Pfaffenwaldring 57,
70569 Stuttgart, Germany}

\begin{abstract}
Quantum information technologies require networks of interacting defect bits.  Colour centres, especially the nitrogen vacancy (NV$^-$) centre in diamond,
represent one promising avenue, toward the realisation of such devices.
The most successful technique for creating NV$^-$ in diamond is ion implantation
followed by annealing.
Privious experiments have shown that shallow nitrogen implantation ($<$10keV)
results in NV$^-$ centres with a yield of $0.01-0.1 \%$.
We investigate the influence of channeling effects during shallow implantation and  statistical diffusion of vacancies
using molecular dynamics (MD) and Monte Carlo (MC) simulation techniques.  Energy barriers for the diffusion process were calculated using density functional theory (DFT).
Our simulations show that 25$\%$ of the implanted nitrogens form an NV centre,
which is in good agreement with our experimental findings. 
\end{abstract}

\keywords{Nitrogen-vacancy, diamond, annealing, implantation, molecular dynamics, vacancy diffusion}

\maketitle
\section{Introduction}

A basic requirement for applications of solid-state defects in quantum information processing (QIP), magnetometry and even for biomarkers
is the accurate placement of impurities in the target material. A promising candidate for these 
applications is the negatively charged nitrogen-vacancy (NV$^-$) centre embedded in diamond. 
The defect consists of a carbon (C) atom which is replaced by a nitrogen (N) together with an adjacent vacancy (V)  
and an additional electron captured from the diamond lattice.\\
\\
Experimentally, ion implantation techniques are used to introduce N and V defects in the carbon matrix.
Upon annealing the diamond above 750K \cite{14} for several hours, the fixed N atoms capture mobile V creating NV centres. Experiments have shown that shallow nitrogen implantations ($<$10keV)
result in a NV$^-$ yield of $0.01-0.1 \%$ \cite{1} while implantation at higher energies result in an NV center yield of 45$\%$\cite{1}. 
To resolve this apparent discrepancy we carried out molecular dynamics investigations of the implantation and annealing process at low energy. \\
\\
Specifically we present simulation results of NV centre creation. 
We study the statistical influence of channeling effects\cite{22} on shallow implantations. 
Further we focus on the statistical diffusion of vacancies in diamond and its impact on the NV centre yield. Our simulations show a significant 
difference in the NV yield compared with earlier experiments. Simulations suggest that 25$\%$ of the implanted nitrogens form an NV centre 
after annealing. Additionally, we compare to experimental results of shallow implantion in the range of $2.5-20$keV. The experimental results are in good agreement with the simulation. 

\section{Method}

Using the ITAP Molecular Dynamics (IMD) Program\cite{imd1,imd2}  
 a diamond system of $5.5\times5.5\times52$nm$^3$ was used for simulations;
the C-C and N-C interaction were represented using a Tersoff potential \cite{2,3,4,5,6} with 
interaction parameters from refs\cite{7,8}. For decreasing C-C or N-C distances 
the potential energy of the Tersoff potential reaches a constant limiting value allowing the atoms to move arbitrarily close to one other. 
Therefore, an additional Ziegler-Biersack-Littmark (ZBL) potential was added \cite{9,10}. For atom separations $r$$<$0.3\r{A}, 
the Tersoff potential is smoothly switched to the ZBL potential. Functions
 for turning-on and turning-off potentials are given by $f_{on} = 1-e^{-12\left(r+0.05\right)^2}$
 and $f_{off} = e^{-12\left(r+0.05\right)^2}$ where $r$ is the distance of the atoms. 
A Nos$\acute{\text{e}}$-Hoover thermostat was used to simulate  
temperature \cite{11,12,13}.\\
\\
In order to bring a system into equilibrium at room temperature, the thermal cycle
shown in figure \ref{fg_ini_Tvstime_6parts} was employed:
\begin{figure}[htb]
\begin{center}
\includegraphics[width=0.9\columnwidth]{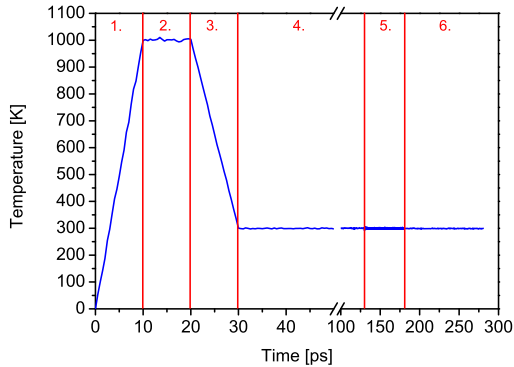}\\
\caption{Schematical initialization plot: Temperature versus time including six simulation steps.}
\label{fg_ini_Tvstime_6parts}
\end{center}
\end{figure}
It consists of five NPT (number of particles, pressure and temperature are conserved) and one final NVE (number of particles, volume and pressure are conserved)
ensemble simulation step. The first four are used to heat the structure up to 1000K, relax it at 1000K, cool it down to room temperature (RT) and to relax the diamond at RT using periodic boundaries in all directions.
In the last NPT step the periodic boundary conditions in the $z$ direction (implantation direction) are removed.
The final NVE simulation equilibrates the system with the ground layer of the diamond fixed in the $z$ direction.
Figure \ref{fg_ini_Tvstime_6parts} shows a plot of the diamond temperature over the simulated time. The six simulation steps are indicated in the plot.\\
\\
All NPT ensemble simulations have been done with a time step of 0.1fs and the NVE equilibration with time steps of 0.01fs. 
The implantation process itself was simulated by adding the impact nitrogen atom on top of the
diamond surface ((100) direction), at a distance $d = 2.51$\r{A} above the surface; this distance is beyond the range of the Tersoff potential. 
The particle is then implanted on a normal incident trajectory with kinetic energy $E_k=4$keV.\\
\\
We analyse the structure after implantation by comparing the damaged structure  
with the initial structure before implantation (reference 
structure). Around each atom position of the reference structure a shell with radius  
$r = 1.54$\r{A} is placed. Counting the number of atoms of the damaged structure within the shell allows
 to analyze the integrity of the diamond lattice (0 atoms=Vacancy; 1=Lattice atom; 2=Interstitial atom and lattice atom;
2$>$Complex structure). We also investigate the infulence of impact atom parameters on the following quantity: 
Depth of the impact atom, displacement in $xy$-direction (straggling) and kinetic energy of the impact atom.\\
\\
To get qualitative insight into our MD simulation we compare Crystal-TRIM (CTRIM)\cite{ctrim} calculations to MD results. 
The CTRIM method uses a ZBL potential for atom interactions and includes an electronic stopping power which is missing in the MD calculations. The results are discussed later in the paper.\\
\\
To investigate the probability for NV centre creation during annealing, we also conducted kinetic Monte-Carlo 
simulations based on the simple hopping frequency of defects:
\begin{eqnarray}
 \Gamma = \omega_{a}e^{-\frac{E_a}{k_BT}}.
\end{eqnarray}
Here $\omega_{a}=10^{13} \text{s}^{-1}$ denotes the attempt frequency with 
which defects attempt to overcome a barrier of energy $E_a$. The values for activation 
energies are shown in table \ref{aenergy}. 
\begin{table}[h]
  \centering
  \begin{tabular}{|c|c|} \hline
      \textbf{Defect}              & \textbf{Migration energy $[$eV$]$}    \\ \hline \hline
           V     \cite{14,20}                  &  2.3                      \\ \hline
           N$_I$ \cite{16}                     &  1.7                      \\ \hline
           C$_I$ \cite{15}                     &  1.5                      \\ \hline
  \end{tabular}
  \caption{Activation energies for vacancy (V), insterstitial nitrogen (N$_I$) and interstitial carbon (C$_I$).}
  \label{aenergy}
\end{table}

\begin{figure}[htb]
\begin{center}
\includegraphics[width=0.6\columnwidth]{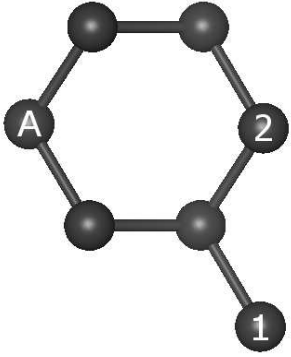}
\caption{Two different atom configurations of N (labeled A) and V (labeled 1 or 2) in the diamond matrix. }
\label{vmd_snap}
\end{center}
\end{figure}

Additionally, DFT simulations using the ABINIT\cite{ABINIT1,ABINIT2,ABINIT3} package have been conducted to investigate a possible uphill potential a vacancy has to pass when approaching a nitrogen atom.
Such a potential would considerably reduce the likelihood of NV centre formation.
For the calculations the projector augmented wave (PAW) method and a $2\times2\times2$ Monkhorst-Pack grid were used for k-point sampling. The exchange correlation potential ($V_{xc}$) was approximated by local density approximation (LDA). The examined structures, including second, third, fourth and sixth next neighbour configurations of an on-site nitrogen and a vacancy,
showed the same total energy within errors. The only exception is a specific configuration with a vacancy on the most distant third next neighbor positions (figure \ref{vmd_snap}). 
With a nitrogen on lattice site A it is energetically more favorable for a vacancy to be at the third next-neigbor site labeled 1 instead of 2.
This is due to symmetry reasons since the lattice is able to relax which decreases the total energy by about 0.7eV.
Except of the above mentioned case no evidence for an uphill potential in vacancy diffusion could be found.

To simulate annealing for implantation close to the diamond surface we considered defect structures of different implantation energies. Since IMD does not take the 
electronic stopping into account, only the 4keV defect structure is simulated by this method. For higher implantation 
energies, defect structures simulated by the Stopping and Range of Ions in Matter (SRIM) \cite{9,19}
($5-50\text{keV}$) are used. The crystal size chosen in the simulations corresponds to
a final concentration of about 
$1\cdot 10^{17} \text{cm}^{-3}$ nitrogen atoms by taking into account the distribution in stopping depth. A simulated time of 2 hours of annealing at 1100K 
decreases the dynamics of vacancies within the crystal such that only a negligible number ($<$5$\%$) are still mobile the end of the simulation.

\section{Results}
\subsection{Implantation}
\begin{figure}[htb]
\begin{center}
\includegraphics[width=0.9\columnwidth]{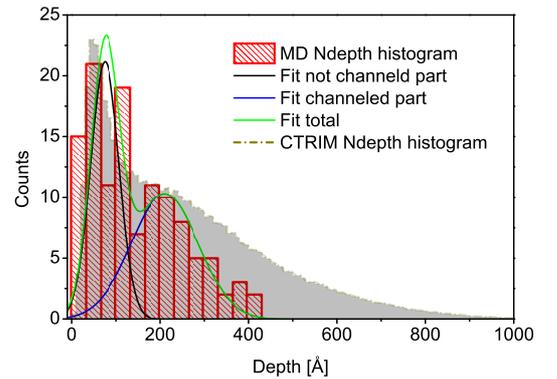}
\caption{Depth distribution of single implanted nitrogen atoms with an incident angle of 0$^{\circ}$ and
with an implantation energy of 4keV; red: MD simulation, grey: CTRIM simulation. The
black and blue curves are Gaussian fits summarized in a cumulative curve (green).}
\label{Ndepth_distr_N_4keV_0grad}
\end{center}
\end{figure}

In recent years, near-surface impurities, in particular NV centres in diamond, have been extensively studied experimentally. Magnetic field imaging\cite{steinert}, 
fluorescence resonance energy transfer (FRET)\cite{tisler} and 
nuclear magnetic resonance (NMR)\cite{staudacher} using NV$^-$ centres in diamond rely on the weak magnetic dipole interaction which decreases with the distance by $1/r^3$. 
Accordingly, sensing applications require reliable production of shallow NV$^-$.\\
Several studies have shown a significant influence of channeling effects on the depth distribution of implanted nitrogens in diamond giving a larger than expected implantation depth\cite{19,ofori-okai}. 
Usual methods such as SRIM, underestimate the defect depth by a factor of two\cite{ofori-okai} as SRIM assumes an amorphous structure for diamond.
We study implantation processes using IMD and CTRIM which take the crystal structure into account.\\
Figure \ref{Ndepth_distr_N_4keV_0grad} shows the depth distribution of 4keV nitrogen implantation along (100) direction. The time demanding MD calculations were repeated 
120 times to provide statistics, in each of which the N starting position was varied
in order to get statistical average values for the depth distribution. The results of MD calculations were compared to CTRIM simulations (Figure 3).

A broad distribution can be extracted from simulation resulting from a combination of two processes, a  ``un-channeled'' - at lower depth,  and a ``channeled'' - at larger depth.
In MD simulations, nitrogens are implanted at a depth of 
 $0-460$\r{A}. The data can be fitted using two Gaussian peaks at $d_{c1} = 75$\r{A} (first
 peak: black curve) and $d_{c2} = 210$\r{A}
(second peak: blue curve). The first Gaussian peak has a FWHM of 77\r{A} whereas the 
second Gaussian peak has a FWHM of 177\r{A}. The number of implanted nitrogen atoms at 
$d_{c1}$ is 21 and at $d_{c2}$ 10. 
In addition the simulations show that $37\%$ of implanted nitrogens end up on a lattice 
site and 63$\%$ end up as interstitials.\\ 
\\
The CTRIM results (figure \ref{Ndepth_distr_N_4keV_0grad}) differ from the MD outcome. Here the number of projectiles was chosen to be $10^6$; 
the CTRIM results are scaled in order to make them comparable with MD results.
The ``un-channeled'' part has a 25$\%$ narrower and the ``channeled'' part a $33\%$ broader distribution than in MD simulation.
The maximum of the ``un-channeled'' peak is shifted by 25\r{A} (33$\%$) towards lower implantation depth whereas the maximum of the ``channeled''
peak is shifted by 20\r{A} (10$\%$) towards higher implantation depth compared to MD.
This might be due to the fact that during implantation a higher number of nitrogens enter the crystal, i.e. in CTRIM one implants into the pre damaged lattice. 
Additionally the empirical potential used in both methods give rise to different results. 
The bond ordered Tersoff potential used in MD describes the interaction between the projectile and the diamond atoms more accurately
than the ZBL potential used in CTRIM. This leads to a higher number of channeled projectiles.

\subsection{Annealing}

\begin{figure}[htb]
\begin{center}
\includegraphics[width=0.9\columnwidth]{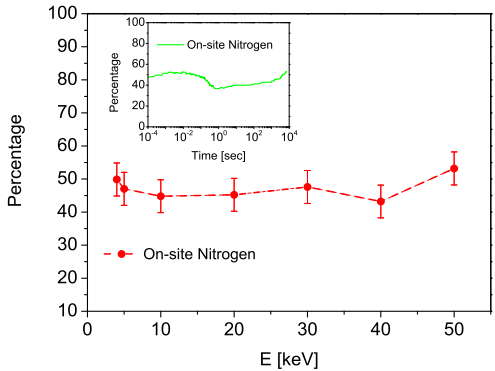}
\caption[Number of on-site nitrogens after annealing depending on the implantation energy.]{Number of on-site nitrogens after annealing depending on the implantation energy. The inset shows the time evolution of on-site nitrogens during annealing after 50 keV implantations with an initial 40$\%$ on-site and 60$\%$ interstitial distribution. The number of on-site nitrogen atoms increases already within the first simulation steps which are not shown in the inset.}
\label{time_evo_onsiteN_mixed}
\end{center}
\end{figure}

After implantation it is unlikely to find a next nearest neighbour pair of a nitrogen atom and a vacancy. An annealing step, is required such that
 vacancies become mobile and diffuse through the lattice.
Several studies on diffusion near to the surface have been done in recent years\cite{Halicioglu_sf,Hu_sf,long_sf,Orwa_sf}. 
Here we model the diffusion process during annealing and investigated its influence on the NV centre yield.\\
The statistical probability for vacancies, created during implantation, to migrate to a next neighbor position of the implanted nitrogen was examined by Monte Carlo simulations.
Our analysis included the most important loss mechanisms for vacancies, such as migration to the surface or 
recombination with interstitials. The processes in which a self-interstitial atom replaces an 
on-site nitrogen at a lattice site and turns the latter into an interstitial nitrogen 
atom \cite{20} was also included. In contrast to on-site nitrogen atoms which hardly move at 1100K, 
interstitial nitrogen atoms are mobile during annealing \cite{20,21}. The initial distribution after 
implantation was assumed to be 40$\%$ on-site nitrogen and 60$\%$ between lattice sites, as suggested by the MD simulations discussed above. 
100 annealing runs with different starting configurations were performed.
Fig. \ref{time_evo_onsiteN_mixed} 
shows the percentage of nitrogen atoms that are found on lattice sites after annealing. 

The number is almost identical for all energies within error bars. 
Hence only about half of the implanted nitrogen atoms are available for NV centre creation in the 
first place. This is in good accordance with the maximum yield found by Pezzagna \textit{et al.} \cite{1} for MeV 
implantations. It suggests that for high energies every implanted nitrogen that finally 
ends up on a lattice site is turned in to a NV$^-$ centre.
\begin{figure}[htb]
\begin{center}
\includegraphics[width=0.9\columnwidth]{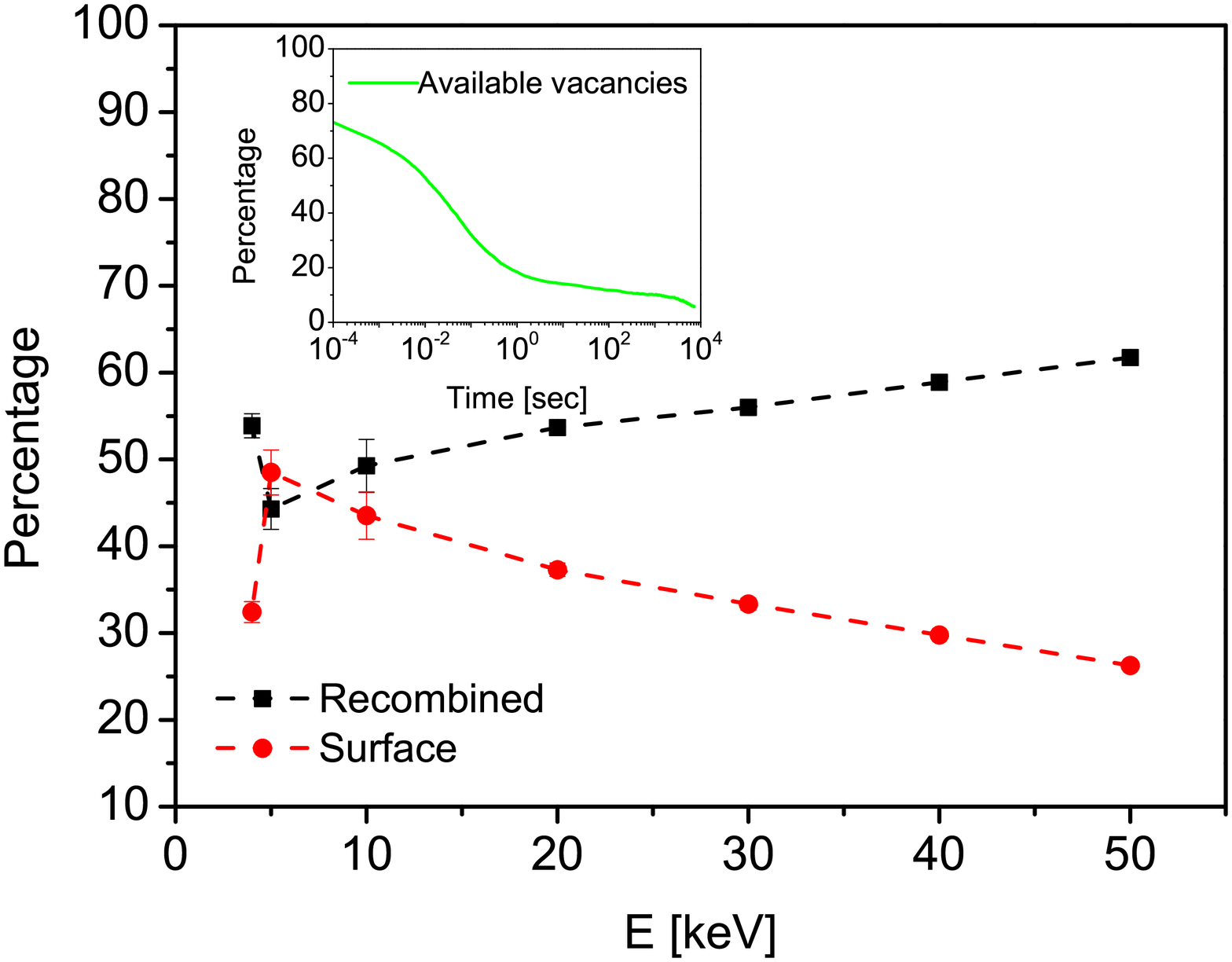}
\caption[Number of vacancies which are lost to the surface or lost due to recombination processes.]{Number of vacancies which are lost to the surface (red) or lost due to recombination processes (black) after 2 hours of simulated annealing. The inset shows the number of vacancies that are still available for NV centre creation over time.}
\label{time_evo_mixed}
\end{center}
\end{figure}
Figure \ref{time_evo_mixed} shows the percentage of initially created vacancies that are lost at the surface or 
due to recombination. By considering intial SRIM input (5-50keV), as expected, at shallow implantations more vacancies (48$\%$) are lost at the 
surface than deep ones inside the lattice (26$\%$). This is because of the short distance to the surface that makes it more likely for vacancies and 
self-interstitials to randomly migrate to the surface where they are lost, instead of recombining. Except for very low implantation energies ($<$20keV) the loss due to 
recombination with interstitial carbon is significantly higher. It increases  from 44$\%$ at 5keV to 62$\%$ at 50keV while the loss at the surface decreases from 49$\%$ to 26$\%$. 
The line crossing in figure \ref{time_evo_mixed} of the vacancies lost 
at the surface and due to recombination defines a threshold; For implantation energies lower than 7keV loss of vacancies is dominated by the surface process, 
for higher implantation energies recombination process and reduces the available vacancies to form NV centres.
The simulations show
that the recombination of vacancies and interstitials happens within the first seconds of 
annealing because of the high mobility of carbon interstitials. Hence the number of vacancies 
that are left to possibly migrate towards a nitrogen atom is less than half the number that is 
suggested by implantation simulations such as SRIM, which does not take recombination into account. 
The results at 4keV implantation energy based on IMD implantation simulations show an even higher 
probability of recombination than the results based on SRIM data. IMD seems to yield defect 
structures with smaller distances between vacancies and the corresponding interstitials which 
therefore are more likely to recombine. The inset in figure \ref{time_evo_mixed} shows the percentage of 
remaining vacancies that are still available for NV centre creation. It can be seen that at the end of the simulation the vacancy 
diffusion within the crystal has almost completely come to an end. On average 90$\%$ of the created vacancies during the implantation process are lost after 2 hours of annealing.\\
Summarising the result of implantation and anneling simulations figure \ref{yield} shows the yield of NV centres of any charge state that is to be expected. 
It is about the same for all investigated implantation energies roughly 25$\%$ of implanted ions are expected to yield a NV center.

\begin{figure}[ht]
\begin{center}
\includegraphics[width=0.9\columnwidth]{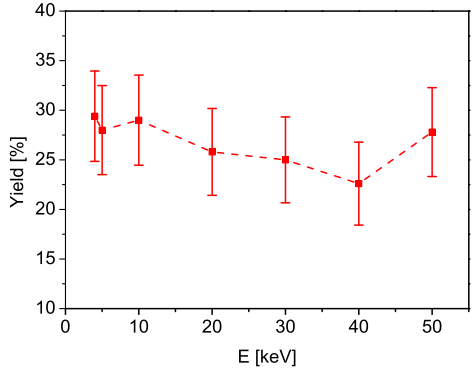}
\caption[Number of NV centres created after implantation and annealing.]{Number of NV centres created after implantation and annealing.}
\label{yield}
\end{center}
\end{figure}

\section{Comparison to experiment}
Implantations experiments were done to experimentally verify the simulation results.\\ 
The characterisation of the shallow NV centres were performed on a custom built confocal microscope at room temperature.
Fluorescence in the 600-900nm wavelength region was collected, allowing both NV$^-$ and NV$^0$ charge states to be observed at the single defect level.
Additionally a flippable mirror was placed in the detection channel which deflected the fluorescence into a spectrometer. 
This enabled the signal of single NV centres originated from $^{15}$N$^+$ to be recorded\cite{rabeau} and to observe their charge states by distinguishing between NV$^-$ and NV$^0$.

\begin{figure}[ht]
\begin{center}
\includegraphics[width=0.49\columnwidth]{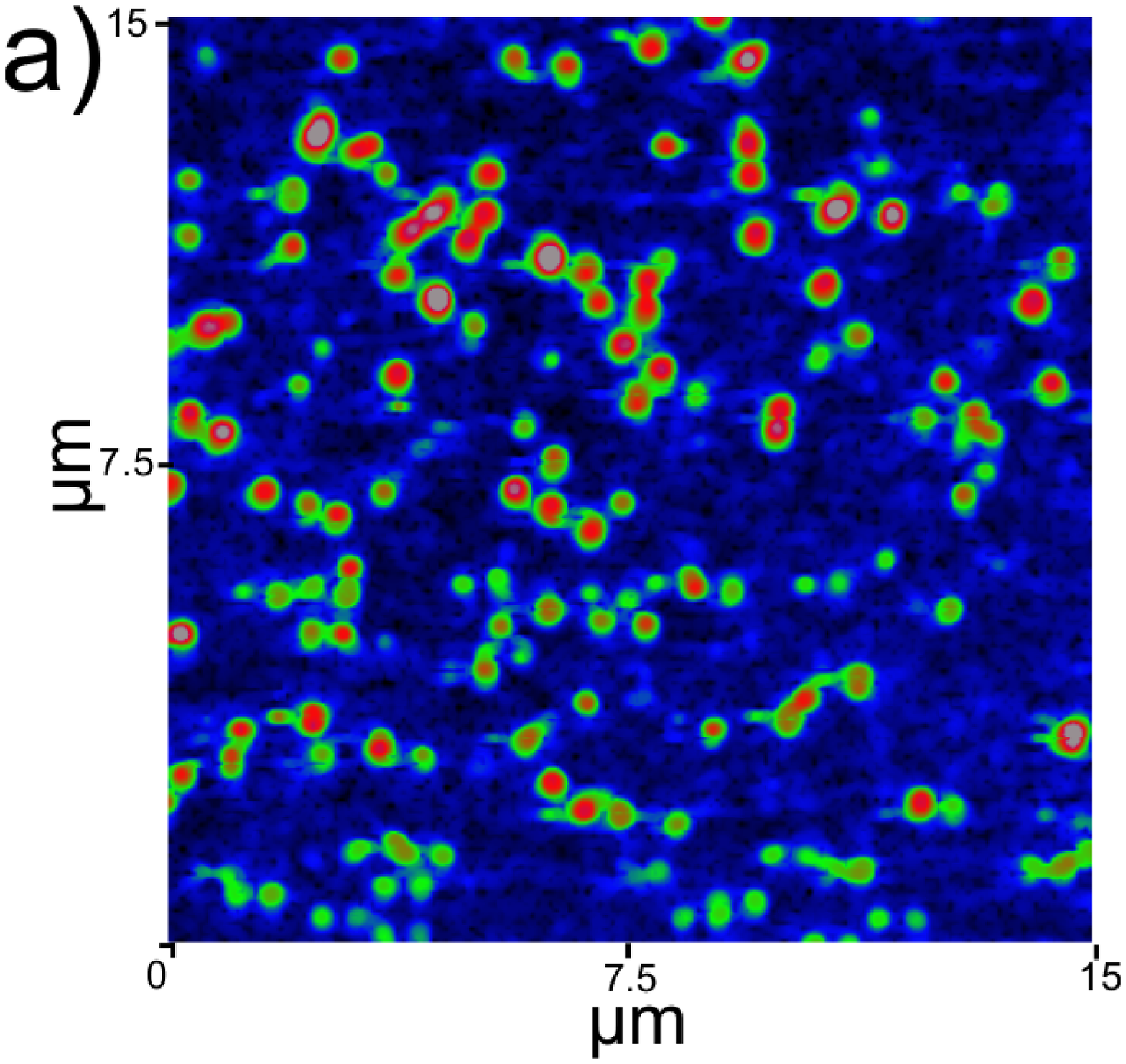}
\includegraphics[width=0.49\columnwidth]{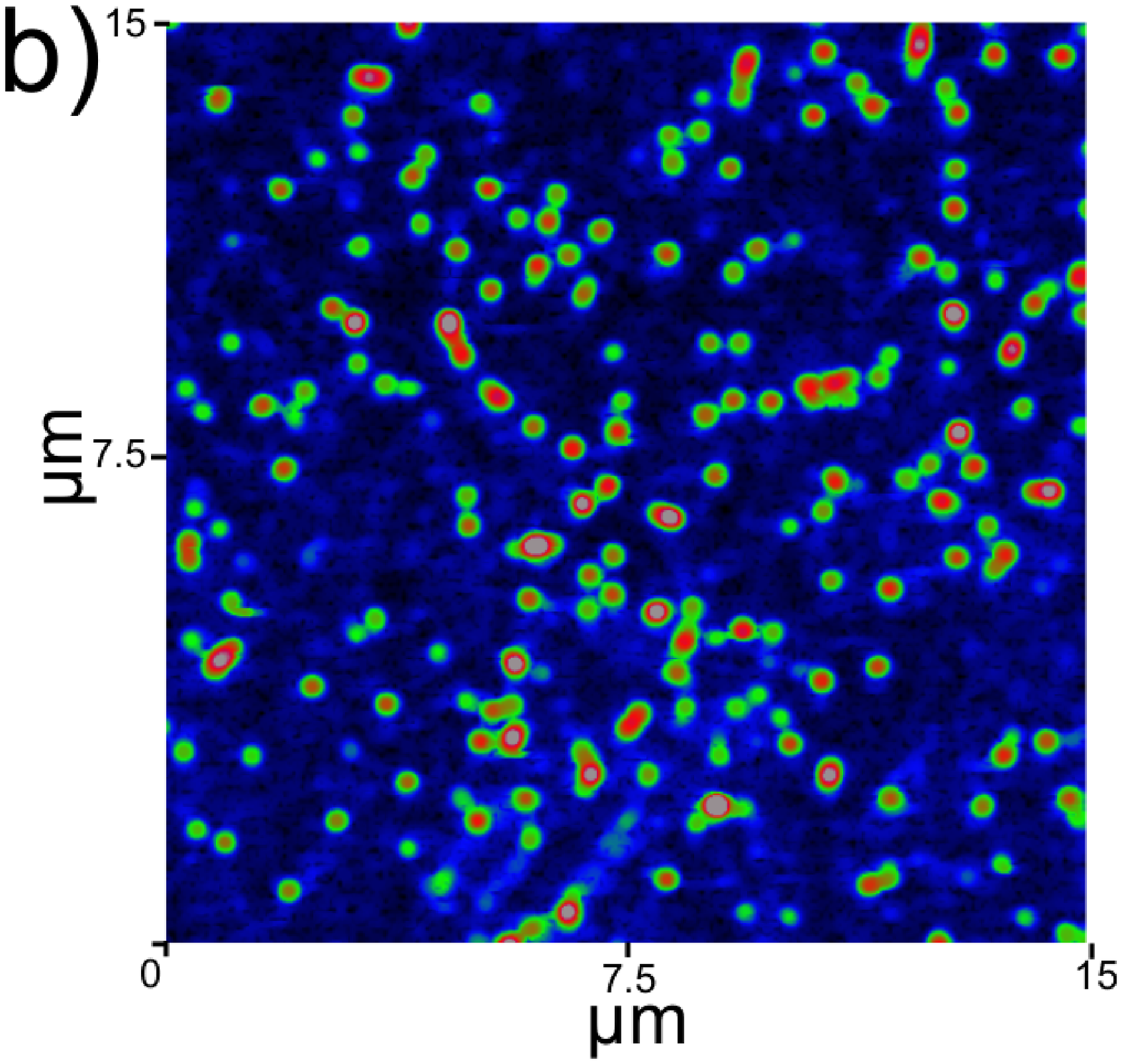}
\includegraphics[width=0.49\columnwidth]{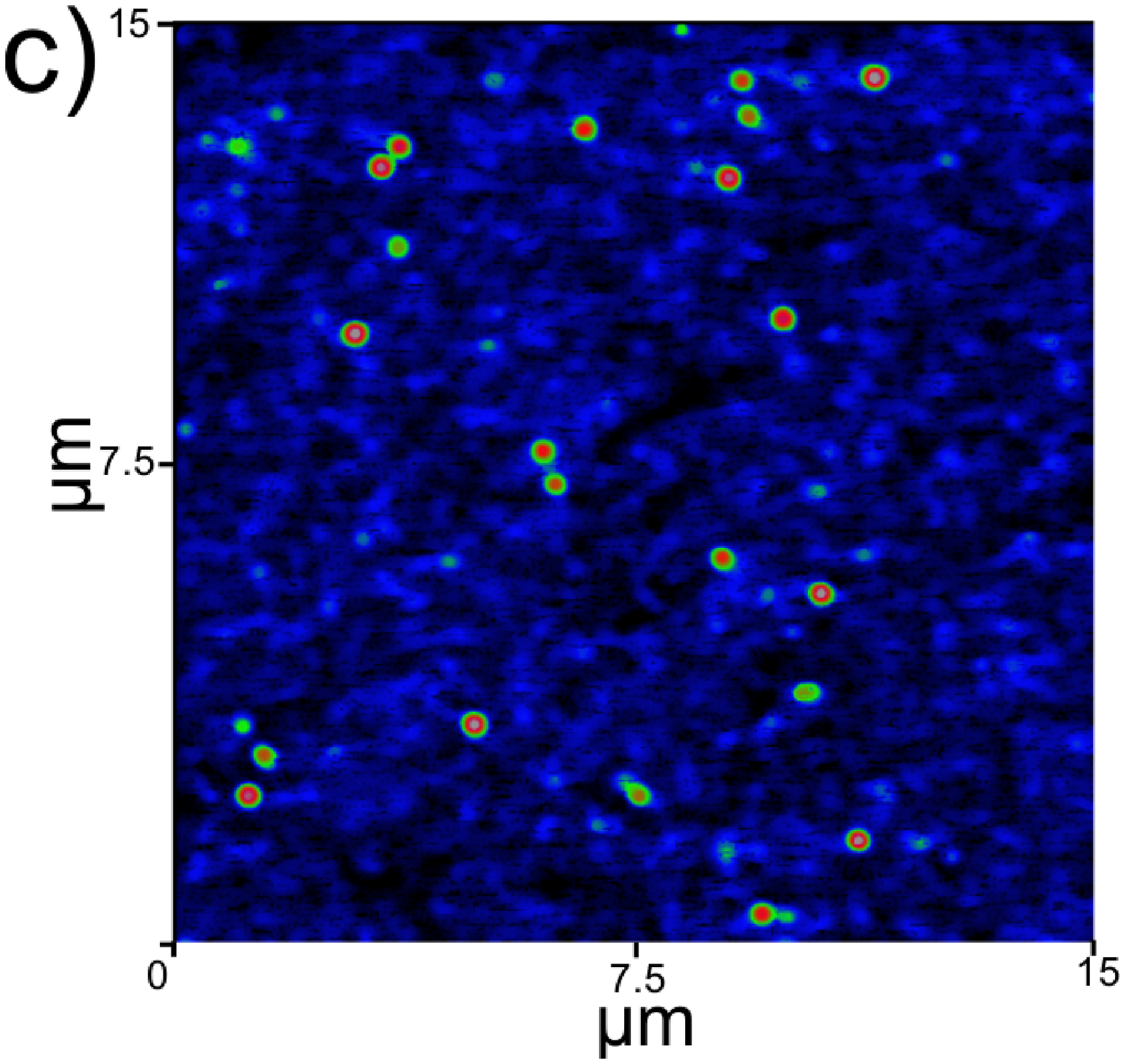}
\caption[Fluorescence images of diamond surfaces.]{Fluorescence images of diamond surfaces for different implantation energies: a) $2.5$keV, b) 5keV, c) 20keV; 
and doses: a)-b) $3\times10^8$ions/cm$^{-2}$, c) $5\times10^7$ions/cm$^{-2}$.
Note, the 20keV implantation was performed with $^{15}$N$_2^+$ whereas the rest was performed with $^{15}$N$^+$. 
The color labelling show the recorded photon counts per second.}
\label{expimages}
\end{center}
\end{figure}

Shallow NV centres were created by implanting low energy nitrogen ions and molecules into ultra pure (nitrogen concentration $<$0.1ppb), $99.999\%$ $^{12}$C enriched diamond. 
$^{15}$N$^+$ ion implantations were performed for $2.5$keV and 5keV implantation energy. $^{15}$N$_2^+$ molecules were accelerated up to 20keV, 
resulting in an implantation energy of 10keV per ion. 
The implantations were done for low doses ($3\times10^8$ions/cm$^2$, $5\times10^7$ions/cm$^2$)
 to minimise implantation damage to the diamond crystal and allow observation and counting of single defects.
The diamond samples were subsequently annealed under high vacuum ($10^{-7}$torr) at a temperature of 1100K for 1 hour. 
High vacuum is essential to avoid etching of the first diamond layers leading to a lower observed yield. 
After annealing the diamonds were boiling in a mixture of sulfuric, nitric and perchloric acids for several hours at 180$^{\circ}$C, 
in order to give a heterogenous oxygen terminated diamond surface. 
This surface treatment is known to allow observation of NV$^-$ centres at low depths. \\
Figure \ref{expimages} shows confocal images of diamond surfaces after the implantation process.
In each image the total number of created NVs, consisting of NV$^-$ and NV$^0$, of an area of $15\mu$m$\times15\mu$m was counted and the yield calculated. 
The results are given in table \ref{expyield}.

\begin{table}[h]
  \centering
  \begin{tabular}{|c|c|c|} \hline 
      \textbf{Implantation energy} & \textbf{Doses}                   & \textbf{Yield}  \\ 
               $[$keV$]$               & $[$ions/cm$^2$$]$                &   [$\%$]        \\ \hline \hline 
           $2.5$                   &                                    & 23                \\ [-1ex]
           $5$                     &  \raisebox{1.5ex}{$3\times10^8$}   & 29                \\ \hline
           $20^*$                  &  $5\times10^7$                     & 21                \\ \hline
        $30^{\substack{* \\ **}}$  &  $2\times10^8$                     & 20                \\ \hline
  \end{tabular}
  \caption{NV yield depending on the implantation energy and the doses. Note, the data labeled with $^*$ are 
performed with $^{15}$N$_2^+$ which share the implantation energy. For all the other implantations $^{15}$N$^+$ is used.
Data labeled with $^{**}$ is taken from Ref. \cite{boris_kalish}}
  \label{expyield}
\end{table}
The data show that the number of created NV centres depends on the implantation energy and on the implantation dose.
The yield of the total number of NV centres increases just slightly for a dose of $3\times10^8$ions/cm$^2$ with implentation energy, resulting in 23-29$\%$. 
No change in yield ($\sim$20$\%$) for higher implantation energies (20keV, 30keV) with similiar implantation dose is observed.

\section{Summary and Discussion}

In this paper we present results using MD, DFT and MC simulation techniques to investigate the NV centre yield after nitrogen implantation  in diamond.
Additionally we show experimental data and compare them to these simulation.
We could demonstrate that during nitrogen implantations into diamond even at low implantation energies (4keV) channeling effects 
occur and have to be taken into account for accurate placement in the carbon matrix. Furthermore, annealing of the implanted diamond 
structures leads to a final NV centre yield of about 25$\%$. This result is significantly higher than the NV$^-$ yield seen in earlier experimental publications 
($0.01-0.1 \%$)\cite{1} with implantation energies $<$10keV. The discrepancy can be explained by the  
charge state of the shallow implanted NV centres or etching of the diamond during the annealing process. Experimentally only the NV$^-$ and the NV$^0$
can be detected. Surface effects or damage structures in the vicinity of the NV centre could change the charge of the NV$^-$; leaving the NV$^+$ state which has not been detected yet. 
Experimentally a maximum NV yield (NV$^-$ and NV$^0$) up to 29$\%$ for implantation energies in a range of $2.5$keV$-$20keV was found. This is in a good agreement with our simulation results. 
At this point we have to mention that we observed a decrease in the yield for an implantation energy of 5keV by a factor of 3 by increasing the dose by one magnitude.
This effect enhances by lowering the implantation energy or increasing the dose. 
An explanation could be found in the damaged region in the vicinity of the implanted nitrogen which does not recover during the annealing process. 
Saada \textit{et. al}\cite{saada} and Sorkin \textit{et. al}\cite{sorkin} investigated 
the annealing behaviour of diamond using MD simulations. 
They observed that a damaged region with 60$\%$ sp$^2$-bonded carbon atoms  
form a graphitic geometry after annealing. 
Graphitisation may render fluorescence of NV centres undetactable, e.g. by changing the charge state to NV$^+$.
For lower implantation dose or higher implantation energy the damage in the vicinity of the NV defect is reduced. 
After the annealing the NV centre is surrounded by a diamond like structure; leaving the NV detactable .
Note, the simulation scheme does not consider charges and therefore give an average value for NV centre creation of all charge states. Introducing additional electrons to the system which can be 
captured by the NV centres to form the NV$^-$ or carbon coimplantations\cite{boris_kalish} may potentially increase the very low experimental yield.\\
Further investigations on the implantation and the annealing process are necessary. The IMD simulation package considers only the nuclear stopping -  a valid approximation for the low energy implantations - whereas the electronic 
stopping is not taken into account. In addition the NV yield depending on the annealing temperature, the intrinsic nitrogen content and the implantation densities 
have yet to be studied. In particular efficient simulations for the implantation with higher energies require simulations of larger systems which are out of reach with current codes and computing powers. Adding precise calucations of the electronic states of defect structures would yield a comprehensive picture of the defect centre and its surrounding necessary for the understanding of the generation process of larger spin arrays.\\
\\
The authors would like to acknowledge financial support by the DFG via the SFB 716 as well as the Forschergruppe FOR 1493. 

\newpage

\bibliography{bibdb}
\bibliographystyle{apsrev-nourl}

\newpage

\end{document}